\documentclass[twocolumn,showpacs,preprintnumbers,prb]{revtex4-1}

\usepackage{graphicx}
\usepackage{dcolumn}
\usepackage{bm}
\usepackage{amsmath}
\usepackage{amssymb}
\usepackage{hyperref}
 \usepackage{epstopdf}

\begin{document}
\title{Optical Hall conductivity of a Floquet Topological Insulator}
\author{Hossein Dehghani}
\author{Aditi Mitra}
\affiliation{Department of Physics, New York University, 4 Washington Place, New York, NY 10003, USA
}
\date{\today}

\begin{abstract}
Results are presented for the optical Hall conductivity of a Floquet topological insulator (FTI) for an ideal
closed quantum system, as well as an open system
in a nonequilibrium steady-state with a reservoir.
The steady-state, even for the open system, is strongly dependent on the topological phase of the FTI, with
certain phases showing a remarkable near-cancellation from pockets of Berry-curvature of opposite signs,
leading to a suppressed low-frequency Hall conductivity, that also shows an anomalous temperature dependence,
by increasing as the temperature of the reservoir
is increased. Such a behavior is in complete contrast to heating, and arises because of a strong modification
of the effective system-reservoir coupling by the laser.
The Berry curvature of the Floquet modes is time-dependent, and its frequency components
are found to control the main features of the high-frequency Hall conductivity.
\end{abstract}

\maketitle

\section{Introduction}
Topological systems are characterized by a bulk-edge correspondence where
geometric properties of the bulk band-structure have a precise connection with
the nature of excitations at the edge when the system is placed in a confined geometry.
For an integer quantum Hall system for example, bulk bands have a non-zero
Chern number $C$, which also equals the number
of chiral edge modes~\cite{TKNN,Bellissard94,Avron94}. This correspondence implies that many bulk measurements are indirect probes
of edge excitations as well.
For example, the dc Hall conductivity at zero temperature, (which is equal to the Hall conductance measured in
a multi-terminal measurement~\cite{Klitzing80}) is
universal $\sigma_{xy}^{\rm dc}= \sigma_{xy}(\omega=0)=Ce^2/h$, and proportional to the number of
edge-states. Moreover, the optical Hall conductivity
$\sigma_{xy}\left(\omega \neq 0\right)$, while non-universal, nevertheless shows signatures of quantum Hall plateaus
as the external magnetic field is varied~\cite{Morimoto09}. In fact an all optical measurement such as Faraday rotation $\Theta_F$
of linearly polarized light of frequency $\omega$ is related to
the optical Hall conductivity as $\Theta_F\left(\omega\right)\sim \alpha c_n\sigma_{xy}(\omega)$
[$\alpha$ being the fine-structure constant, and $c_n$ being a
material dependent parameter such as the refractive index]~\cite{Wallace82}, and has been used as an alternate probe of quantum Hall
physics, both in semiconductor heterostructures~\cite{Ikebe10}, and graphene~\cite{Crassee11,Shimano13}.

Chern insulators are topological insulators (TIs) which show quantum Hall physics in
the absence of a magnetic field, where time-reversal symmetry is broken by introducing complex
hopping amplitudes~\cite{Haldane88}. This can be achieved by the application of a circularly polarized
laser~\cite{Oka09,Inoue10,Kitagawa10,Lindner11,Fertig11,Podolsky13}, where TIs arising out of such time-periodic
perturbations are referred to as Floquet TIs (FTIs)~\cite{Lindner11}.
The field of FTIs has grown in recent years because of
several experimental realizations ranging from periodically shaken lattices of cold-atomic gases~\cite{Esslinger14}, to
graphene~\cite{Karch10,Karch11} and Dirac fermions on the surface of 3D TIs~\cite{Gedik13} under external irradiation,
and also arrays of twisted
photonic waveguides~\cite{Segev13}. In fact FTIs are extremely rich, showing different topological phases
as the amplitude, frequency, and polarization of the periodic drive is varied~\cite{Rudner13,Kundu14,Carpentier14,Dehghani15a}.
\begin{figure}
\includegraphics[height=9cm,width=9cm,keepaspectratio]{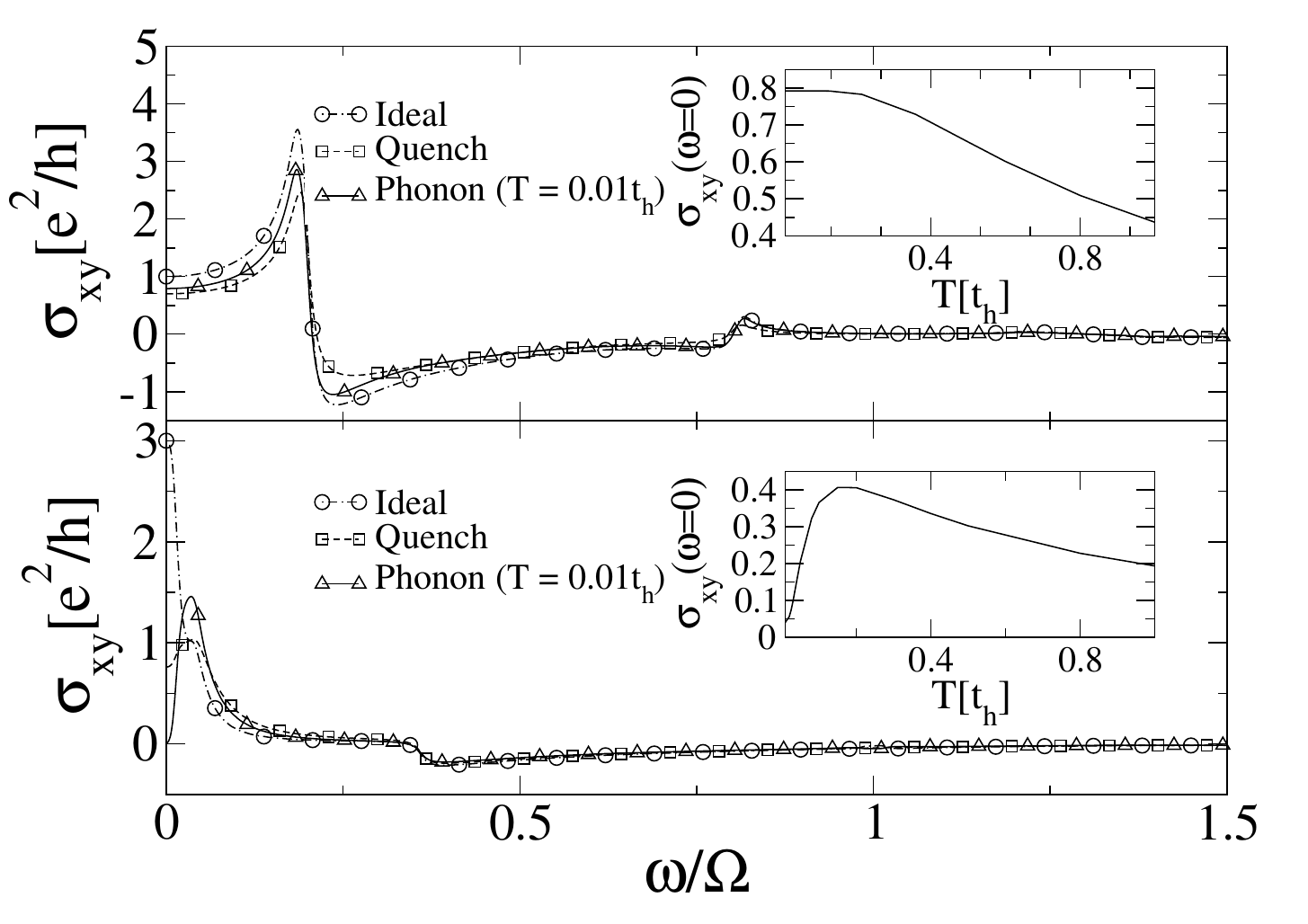}\\
\caption{Optical Hall conductivity (main panel) with temperature dependence of the dc Hall conductivity (inset) for laser frequency $\Omega = 5.0t_h$.
The laser amplitude and Chern number are upper-panel: $A_0a = 1.5, C=1$, lower-panel: $A_0a = 0.5, C=3$ .
}
\label{figOm5}
\end{figure}

\begin{figure}
\centering
\includegraphics[height=8cm,width=8cm,keepaspectratio]{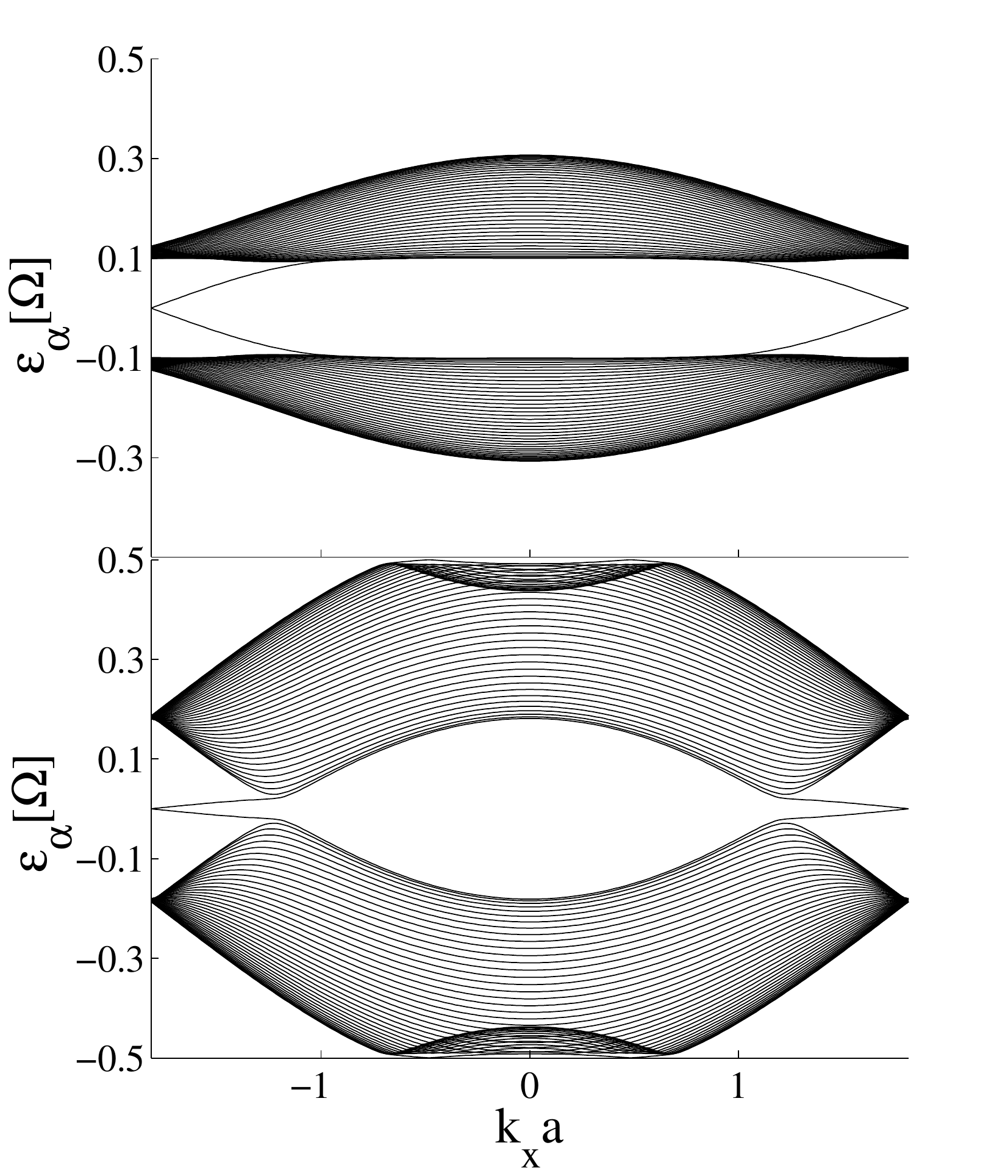}
\caption{Quasi-energies in a cylindrical geometry highlighting edge states. The laser frequency is
$\Omega=5.0t_h$, while the laser amplitude and Chern number are (upper panel) $A_0a=1.5, C=1$  and (lower panel) $A_0a=0.5,C=3$.
The latter shows edge-states both at the center and the edges of the FBZ.
}
\label{figedg1}
\end{figure}

\begin{figure}
\centering
\includegraphics[height=10cm,width=10cm,keepaspectratio]{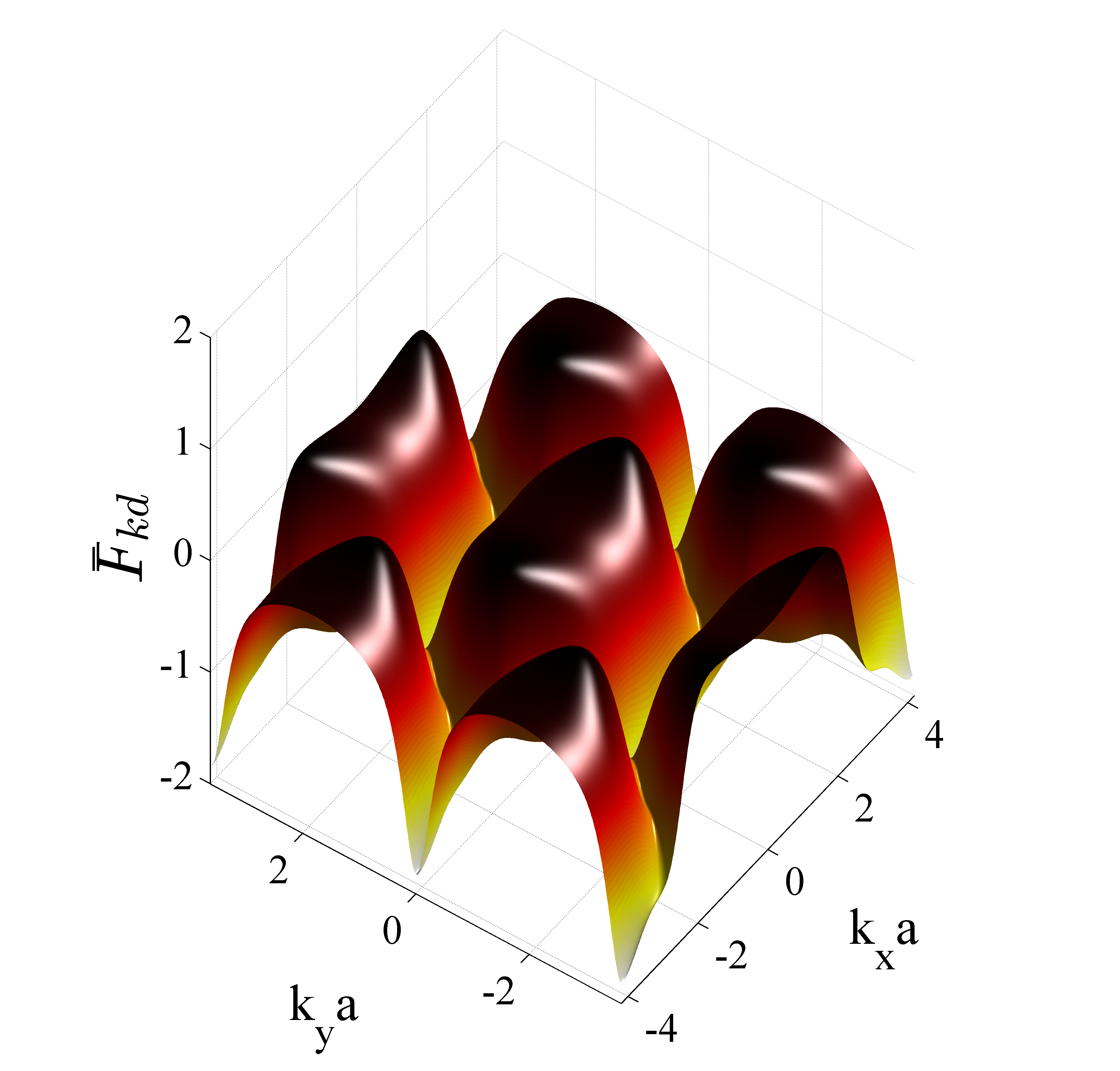}
\caption{(color online) Time-averaged Berry curvature over 4 BZs for $\Omega=5t_h,A_0a=1.5$ corresponding to
a Chern number of $C=1$. The structure is rather smooth, moreover in the Fourier decomposition of
$F_{kd}(t)$, two harmonics ($F_k^{m=0,1}$) are dominant.
}
\label{figFkA1p5}
\end{figure}

\begin{figure}
\centering
\includegraphics[height=10cm,width=10cm,keepaspectratio]{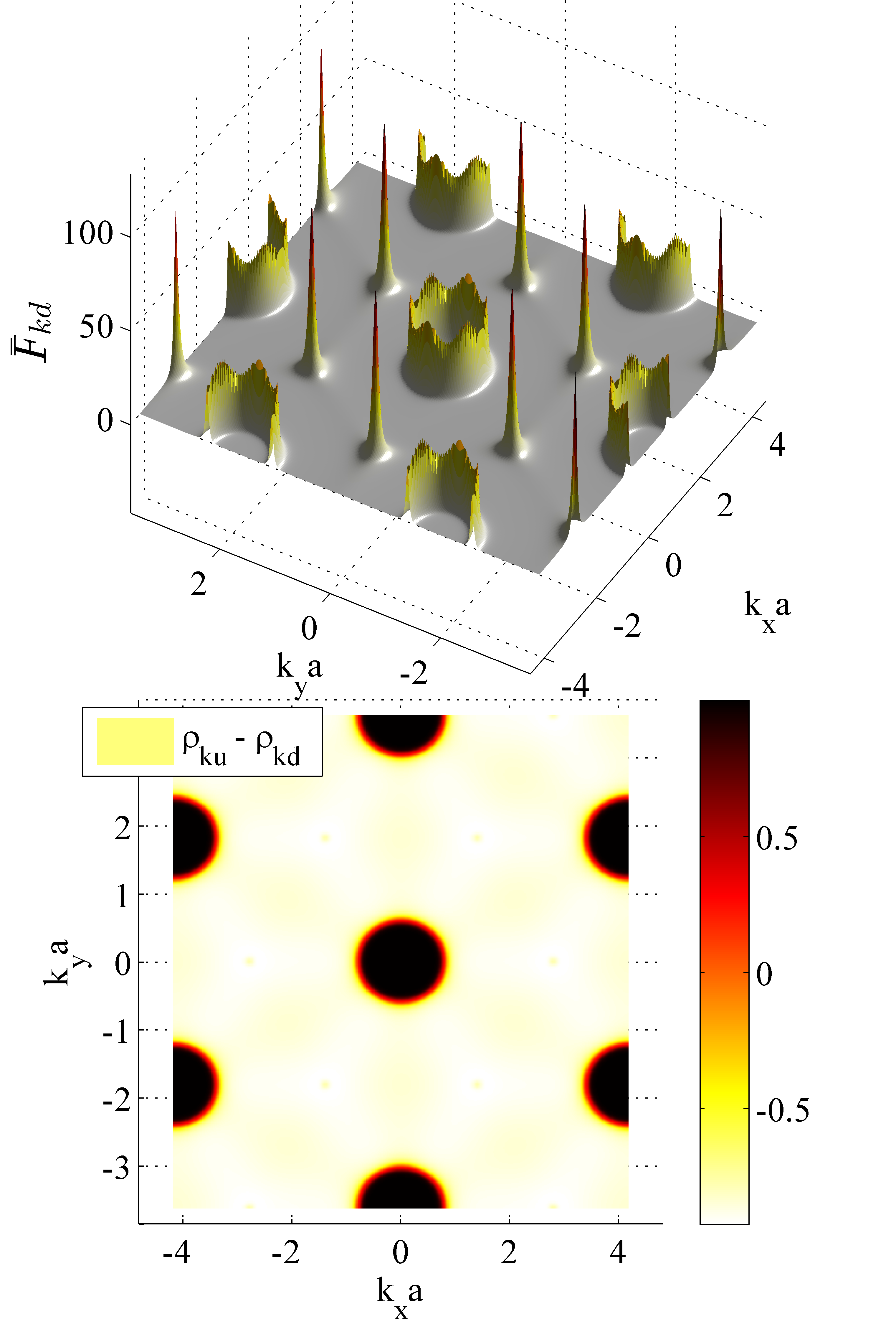}
\caption{(color online) Upper panel: Time-averaged Berry curvature over 4 BZs for $\Omega=5.0t_h,A_0a=0.5, C=3$.
Lower panel: Contour plot for the steady-state population difference for reservoir temperature $T=0.01t_h$.
Note the almost complete population inversion between regions around the
Dirac points and the circles.
}
\label{FkOm5}
\end{figure}

In this work we present results for the optical Hall conductivity of FTIs by accounting for the fact that
these systems are far out of equilibrium. To this end, the precise relaxation mechanisms need to be specified
as they sensitively affect results. Thus we present results for two rather different cases,
one where the system is closed and the laser
is switched on as a quench, while the second is when the system is
coupled to an ideal reservoir, and so the steady-state loses memory of its initial
state. Yet being a driven dissipative system, the steady-state is far from thermal,
resulting in some unusual behavior for the Hall conductivity, which is sensitive to the topological
phase, and cannot be interpreted as simply ``heating''. In fact we show that the laser parameters, and in particular the topological
phase, strongly affects the effective system-reservoir coupling, and therefore the steady-state, and its Hall response.

The advantage of the optical Hall conductivity is that it can be measured both
in traditional multi-terminal measurements~\cite{Dattabook,Torres14,Barnea15}, as well as completely optically via the Faraday rotation, the latter being
more suitable for present day experiments. In fact even the dc limit can be studied without leads, in optical
lattices, by monitoring the transverse drift in time-of-flight measurements~\cite{Esslinger14}.
Moreover while it is the dc ($\omega \rightarrow 0$) limit of the Hall conductivity, combined with
an ideal electron distribution function,
that is related to a topological quantity, namely the Chern number of the Floquet bands~\cite{Oka10,Dehghani15a},
we show that even at non-zero frequencies, some
features of the topological nature of the system survive. In particular, the Berry-curvature of Floquet bands is time-dependent,
and we show that various frequency components of this Berry-curvature can be directly probed in the optical Hall
conductivity.

The paper is organized as follows. The model is introduced in Section~\ref{model} and the derivation of the Floquet-Master equation for the open
system is outlined (with some details relegated to Appendix~\ref{ap1}). In Section~\ref{Kubo}
the optical Hall conductivity is derived using a linear-response approach, while in Section~\ref{results} results for the optical Hall conductivity
are presented.
Finally in Section~\ref{conclu} we present our conclusions.

\section{Model}\label{model}
Our model is
graphene irradiated by a circularly polarized laser, and also coupled to a phonon bath. The Hamiltonian is,
$H = H_{\rm el} + H_{\rm ph} + H_c$
where (setting $\hbar=1$) $H_{\rm el}$ is the electronic part,
\begin{eqnarray}
&&H_{\rm el}\!\!=-t_h\sum_k\begin{pmatrix}c_{kA}^{\dagger}&c_{kB}^{\dagger}\end{pmatrix}\!\!
\begin{pmatrix}0&h_{k}^{AB}(t)\\\left[h_{k}^{AB}(t)\right]^*&0\end{pmatrix}
\!\!\begin{pmatrix}c_{kA}\\c_{kB}\end{pmatrix}
\end{eqnarray}
where $h^{AB}_k(t) = \sum_{i=1,2,3}e^{ia(\vec{k} + \vec{A}(t))\cdot\vec{\delta}_i}$,
$\vec{\delta}_i$ are the nearest-neighbor unit-vectors of graphene, $a$ is the
lattice spacing, and $\vec{A}$ is the
circularly polarized laser of amplitude $A_0$ and frequency $\Omega$,
$A_x(t) = \theta(t)A_0 \cos\Omega t; A_y(t) = - \theta(t)A_0\sin\Omega t$, which we
assume has been suddenly switched on at time $t=0$.
We consider dissipation arising due to 2D phonons
$H_{\rm ph}=\sum_{q,i=x,y}\left[\omega_{qi}b_{qi}^{\dagger}b_{qi}\right]$ which are coupled to the electrons as
\begin{eqnarray}
H_c = \sum_{{k}q\sigma,\sigma'=A,B}
c_{k+q\sigma}^{\dagger}\vec{M}_{\rm el-ph}(k,q)\cdot\vec{\sigma}_{\sigma \sigma'}c_{{k}\sigma'}
\end{eqnarray}
$\sigma$ is a pseudospin label denoting the $A,B$ sub-lattices, and the
electron-phonon coupling $\vec{M}_{\rm el-ph}(k,q)
= \left[\lambda_{x,kq}\left(b_{x,-q}^{\dagger}+b_{x,q}\right), \lambda_{y,kq}\left(b_{y,-q}^{\dagger}+b_{y,q}\right)
\right]$
is off-diagonal in pseudo-spin space~\cite{Ando02,Ando06}. While we will consider only inter-(quasi)-energy-band transitions
driven by long-wavelength optical phonons, our result showing that
the effective matrix-elements of the electron-phonon coupling is modified by the laser, leading to nonequilibrium steady-states
that depend strongly on the topological phase of the FTI, is rather general, and will carry over to other kinds of system-reservoir
couplings.

We assume that initially the system is in the ground
state of graphene $|\Psi(t=0)\rangle=\prod_k|\psi_{\rm in,k}\rangle$, and we time-evolve
the system according to $H$. For the closed system ($H_c=0$), this time-evolution can
be studied exactly, while in the presence of phonons, we solve the
problem by assuming a weak electron-phonon coupling, and a very fast bath
so that a Floquet-Markov approximation can be made~\cite{Hanggi2005,Kohn01,Kohn09,Dehghani14,Dehghani15a,Suppmat}.

Let us denote $W(t)$ to be the full density matrix obeying
$dW(t)/dt = -i \left[H,W(t)\right]$, which
in the interaction representation becomes,
$W_{I}(t) = e^{i H_{\rm ph} t}U^{\dagger}_{\rm el}(t,0)W(t)U_{\rm el}(t,0)e^{- i H_{\rm ph} t}$,
where $U_{\rm el}(t,t')=\prod_kU_k(t,t')$ is the time-evolution operator
for the electrons under a periodic drive, and
$U_k(t,t_0)=\sum_{\alpha=u,d}e^{-i\epsilon_{k\alpha}(t-t_0)}|\phi_{k\alpha}(t)\rangle\langle\phi_{k\alpha}(t_0)|$
with $|\phi_{k\alpha}(t)\rangle,\epsilon_{k\alpha}$ being the quasi-modes and quasi-energies that obey
$\biggl[H_{\rm el}-i\partial_t\biggr]|\phi_{k\alpha}\rangle = \epsilon_{k\alpha}|\phi_{k\alpha}\rangle$. We
restrict the quasi-energies to a Floquet Brillouin zone (FBZ),
$-\Omega/2<\epsilon_{k\alpha}<\Omega/2$, where the two quasi-energy levels are labeled $\epsilon_{ku}>\epsilon_{kd}$. The electron reduced
density matrix, obtained from tracing over the phonons,
$W^{\rm el} = {\rm Tr}_{\rm ph}W$, has the following form
in the interaction representation,
$W^{\rm el}_I(t)  =\prod_k \sum_{\alpha\beta}\rho_{k,\alpha \beta}(t) |\phi_{k\alpha}(t)\rangle\langle\phi_{k\beta}(t)|$.
For a quench switch-on, the physically relevant initial condition is given by the overlap between the initial
state and the Floquet modes at $t=0$,
$\rho_{k,\alpha\beta}(t=0)=\rho_{k,\alpha\beta}^{\rm quench}= \langle \phi_{k\alpha}(0)|\psi_{{\rm in},k} \rangle\langle\psi_{{\rm in},k}
| \phi_{k\beta}(0)\rangle$.
Under assumptions that the bath is Markovian,
always stays in thermal equilibrium at temperature $T$, and that the electron-reservoir
coupling is small in comparison to all quasi-energy level spacings~\cite{Kohn09},
the density matrix obeys the equation of motion,
$\dot{\rho}_{k,\alpha\alpha}(t)
=-\sum_{\beta=u,d}L^k_{\alpha\alpha;\beta\beta}
\rho_{k,\beta\beta}(t)$, where
$L^k_{\alpha\alpha,\beta\beta}$ are the scattering rates.
The steady-state solution is: 
\begin{eqnarray}
&&\rho_{k,\alpha\alpha}(t=\infty)=\rho_{k\alpha}\\
&&\rho_{ku}=\frac{|L^k_{uu,dd}|}{\biggl(|L^k_{uu,dd}|+|L^k_{uu,uu}|\biggr)},\rho_{kd}=1-\rho_{ku}
\end{eqnarray}
where 
the scattering rates for a constant electron-phonon coupling ($\lambda_{x,qk}=\lambda_{y,qk}=\lambda$) with a broad phonon
density of states $D_{\rm ph}$ are,
\begin{widetext}
\begin{eqnarray}
&&L^k_{uu,uu} = \sum_{n=\rm int}M^n_k\biggl[\theta(-\epsilon_{kd}+\epsilon_{ku}+n\Omega)\bar{N}(-\epsilon_{kd}+\epsilon_{ku}+n\Omega)
+\theta(\epsilon_{kd}-\epsilon_{ku}-n\Omega)N(\epsilon_{kd}-\epsilon_{ku}-n\Omega)\biggr]\label{uuuu}\\
&&-L^k_{uu,dd} = \sum_{n=\rm int}M_k^n\biggl[\theta(\epsilon_{kd}-\epsilon_{ku}-n\Omega)\bar{N}(\epsilon_{kd}-\epsilon_{ku}-n\Omega)
+\theta(-\epsilon_{kd}+\epsilon_{ku}+n\Omega)N(-\epsilon_{kd}+\epsilon_{ku}+n\Omega)\biggr]\label{uudd}
\end{eqnarray}
\end{widetext}
Above $\theta(x)$ is the step-function, $N(x) = 1/(e^{x/T}-1)$ is the Bose function, $\bar{N}=1+N$. The matrix elements for
scattering of electrons between the quasi-energy levels $\epsilon_{kd}\leftrightarrow \epsilon_{ku}+n\Omega$ by phonon absorption or emission are
$M^n_k= 2D_{\rm ph}\lambda^2\sum_{\sigma}\left(m^{n}_{\sigma k,ud}m^{-n}_{\bar{\sigma}k,du}\right)$,
where $\langle\phi_{k\alpha}(t)|c_{k\sigma}^{\dagger}c_{k\bar{\sigma}}|\phi_{k\beta}(t)\rangle=\sum_{n={\rm int}}e^{in\Omega t}m^n_{\sigma k,\alpha\beta}$.

\section{Derivation of the optical Hall conductvity}\label{Kubo}
We will now explore the optical Hall conductivity $\sigma_{xy}$ of the open and closed system,
where for the latter $\rho_{k\alpha}=\rho^{\rm quench}_{k,\alpha\alpha}$.
$\sigma_{ij}$ is a response to a weak perturbation applied over and
above the laser, and is computed using linear-response theory~\cite{Oka10,Dehghani15a}.  In particular
the current in the direction $\vec{i}$, in response to a weak electric field $\vec{E}(t)$, is $j_i(t) =\int dt'\sigma_{ij}(t,t')E_j(t')$.
In general $\sigma_{ij}(t,t')$ is not time translationally invariant due to the
time-periodic perturbation. We present results after time-averaging $t+t'$ over a laser cycle $T_{\Omega}=2\pi/\Omega$, and
then Fourier transforming with respect to the time difference $t-t'$. Before presenting explicit expressions for the
$\sigma_{xy}$, note that due to
the time-dependence of the Floquet modes, the Berry curvature
\begin{eqnarray}
F_{kd}(t)=2{\rm Im}\biggl[\langle \partial_y\phi_{kd}(t)|\partial_x\phi_{kd}(t)\rangle\biggr]
\end{eqnarray}
is time-dependent. Yet its integral over the BZ
is time-independent and gives the Chern number  
\begin{eqnarray}
C = \frac{1}{2\pi}\int_{\rm BZ} d^2kF_{kd}(t)
\end{eqnarray}
Defining
the Fourier transform of the Berry ``vector potential'',
\begin{eqnarray}
A^m_{\beta i \alpha}=\frac{1}{T_{\Omega}}\int_0^{T_{\Omega}} dt e^{-im\Omega t}\langle\phi_{k\beta}(t)|\frac{\partial }{\partial {k}_i}\phi_{k\alpha}(t)\rangle
\end{eqnarray}
a natural object that
appears in the optical Hall conductivity is,
\begin{eqnarray}
F^m_{k}=i\biggl[ A^{-m}_{u x d}A^{m}_{d y u}-A^{-m}_{u y d} A^{m}_{d x u}\biggr]
\end{eqnarray}
which  can be thought of as a Fourier decomposition of $F_{kd}(t)$ and is such
that $\sum_{m={\rm int}}F^m_{k}$ is the Berry-curvature time-averaged over one cycle of the laser,
\begin{eqnarray}
&&\sum_{m={\rm int}}F^m_{k}=\overline{F}_{kd}\nonumber\\
&&=\frac{2}{T_{\Omega}}\int_0^{T_{\Omega}}\!\!dt{\rm Im}\biggl[\langle \partial_y\phi_{kd}(t)|
\partial_x\phi_{kd}(t)\rangle\biggr]
\end{eqnarray}
We find that the optical Hall conductivity at steady-state may be written as,
\begin{eqnarray}
\sigma_{xy}\left(\omega\right)=\sum_{m={\rm int}}\sigma_{xy}^m(\omega)
\end{eqnarray}
where $\sigma_{xy}^m$ depend on $F_k^m$ as ($\delta =0^+$),
\begin{eqnarray}
&&\sigma_{xy}^m(\omega) = -\frac{e^2}{\left(2\pi h\right)}
\int d^2k\biggl[\epsilon_{ku}-\epsilon_{kd}-m\Omega\biggr]^2F_k^m\nonumber\\
&&\times \frac{\biggl[\omega^2-\left(\epsilon_{ku}-\epsilon_{kd}-m\Omega\right)^2-2i\delta\omega\biggr]}
{\biggl[\omega^2 -\left(\epsilon_{ku}-\epsilon_{kd}-m\Omega\right)^2\biggr]^2 + 4\omega^2\delta^2}\!\!
\biggl[\rho_{kd}-\rho_{ku}\biggr]
\end{eqnarray}
In the low frequency (dc) limit, this reduces to~\cite{Dehghani15a},
\begin{eqnarray}
\sigma_{xy}(\omega=0)= \frac{e^2}{2\pi h}\int_{BZ}d^2k \overline{F}_{kd}\left[\rho_{kd}-\rho_{ku}\right]
\end{eqnarray}
The optical Hall conductivity depends on the occupation probabilities
$\rho_{k\alpha=u,d}$
and we refer to the ``ideal'' limit as one where only one Floquet band is fully
occupied $|\rho_{kd}-\rho_{ku}|=1$, so that the dc Hall conductivity is
$\sigma_{xy}^{\rm ideal}(\omega=0) = C e^2/h$.
We will show below that while $\sum_mF^m_k$ controls the low frequency Hall conductivity, the individual $F^m_k$ control the
high-frequency Hall conductivity which show enhancement (side-bands) in the vicinity of
$\omega\sim |\epsilon_{ku}-\epsilon_{kd}-m\Omega|$ where $k$ are those points in the BZ where $F_k^m$ are peaked.
We will present results for ${\rm Re}\left[\sigma_{xy}(\omega)\right]$, and choose a disorder broadening
$\delta =\Omega/100$.

\section{Results}\label{results}

The system shows a series of topological phase transitions
as the laser amplitude or frequency is varied, with the topological transitions involving level
crossings at the center and/or boundaries of the FBZ~\cite{Kundu14}. Moreover while topological phases with
non-zero Chern number are possible both for laser frequencies off-resonant ($\Omega > 6t_h$) and resonant $(\Omega <6t_h)$
with the electronic states, the topological phases may be rather different for these two cases,
as for the latter, the laser can create an effective band inversion~\cite{Lindner11},
strongly modifying the Berry curvature. We find that the inelastic matrix-elements $M_k^n$ also depend
on the laser parameters, by being strongly peaked
around $n= n_{0}$, where $n_0$ depends on the
resonance condition ({\sl i.e.}, whether the laser is off-resonant, or single-photon or two-photon {\rm etc}
processes dominate).

In particular, in the limit of
large (off-resonant) laser frequency, and small laser amplitude ($A_0at_h/\Omega\ll 1$), the $M_k^{n=0}$ term dominates, and the
resultant distribution is very similar to the conventional one $\rho_{ku}\rightarrow 1/\biggl[\exp{(\epsilon_{ku}-\epsilon_{kd})/T}+1\biggr]$,
and thus for this case the reservoir "cools" the system, giving a Hall response that approaches the ideal limit as the reservoir temperature
is lowered~\cite{Dehghani15a}. In contrast, as discussed below, topological phases where the laser frequency is smaller than the band-width can
give rise to dominant matrix elements ($M_k^{n_0}$) where $n_0$ are strongly quasi-momentum dependent, resulting in unusual Hall response.

Fig.~\ref{figOm5} shows $\sigma_{xy}$ along with its temperature dependence, for laser frequency $\Omega=5.0t_h$ and
two different amplitudes $A_0 a=1.5, 0.5$, corresponding to two
different topological phases $C=1,3$ respectively. The quasi-energy spectra for these two phases in a cylindrical geometry
highlighting edge states is shown in Fig.~\ref{figOm5}, while the Berry-curvature of the phase with $A_0 a=1.5, C=1$ is shown in Fig.~\ref{figFkA1p5},
and that of the phase with $A_0a=0.5,C=3$ is shown in Fig.~\ref{FkOm5}.

One finds that the Chern insulator for $A_0a=1.5, C=1$ is a conventional one ($n_0=0$), with
the bath effectively cooling the system in that the Hall conductivity increases as the temperature of
the reservoir is lowered.
In contrast the case of $A_0a=0.5,C=3$ is quite different because the low temperature Hall conductivity
is small, almost zero, and has an anomalous temperature dependence in that it actually
increases as the reservoir temperature is increased. In fact the entire region of
$\Omega=5 t_h, 0.01\lesssim A_0a<1,C=3$ constitutes the same
topological phase, and shares this behavior of low dc Hall conductivity, and non-monotonic temperature
dependence.

Fig.~\ref{FkOm5} shows the time-averaged Berry
curvature and the population imbalance
for the above phase ($A_0a=0.5, C=3$).
Besides the characteristic peaks at the Dirac points that one expects in Chern insulators described by the Haldane model, 
$\bar{F}_{kd}$ for these laser parameters also shows
sharp circular rings. While the Dirac points contribute to a Chern number of $1/2$, the
circular rings give a Chern number of $2$, so that in total $C=3$ (since the BZ contains two Dirac points and a ring).
The steady-state distribution function for this case shows that there is effectively a population inversion between
the regions around the Dirac points where the electrons are primarily in the
"down" level, and the regions within the circles, where the electrons are primarily in the ``up" level. This population inversion
arises due to the structure of the the matrix elements $M_k^{n_0}$
where in the vicinity of the Dirac points $M_k^{0}$ is dominant, while within the circles the $M_k^{n=0}\simeq 0$,
while $M_k^{n=-1,-2}$ are dominant. This implies that at low temperatures where phonon absorption is suppressed,
the system can only relax via phonon emission from $\epsilon_{kd}+\Omega \rightarrow \epsilon_{ku}$ within the circles, whereas in the
regions around the Dirac cones, the phonon emission processes cause relaxation from $\epsilon_{ku}\rightarrow \epsilon_{kd}$.
Since the Hall conductivity involves integrating over the entire BZ,
this results in a low Hall conductivity.
\begin{figure}
\centering
\includegraphics[height=9cm,width=9cm,keepaspectratio]{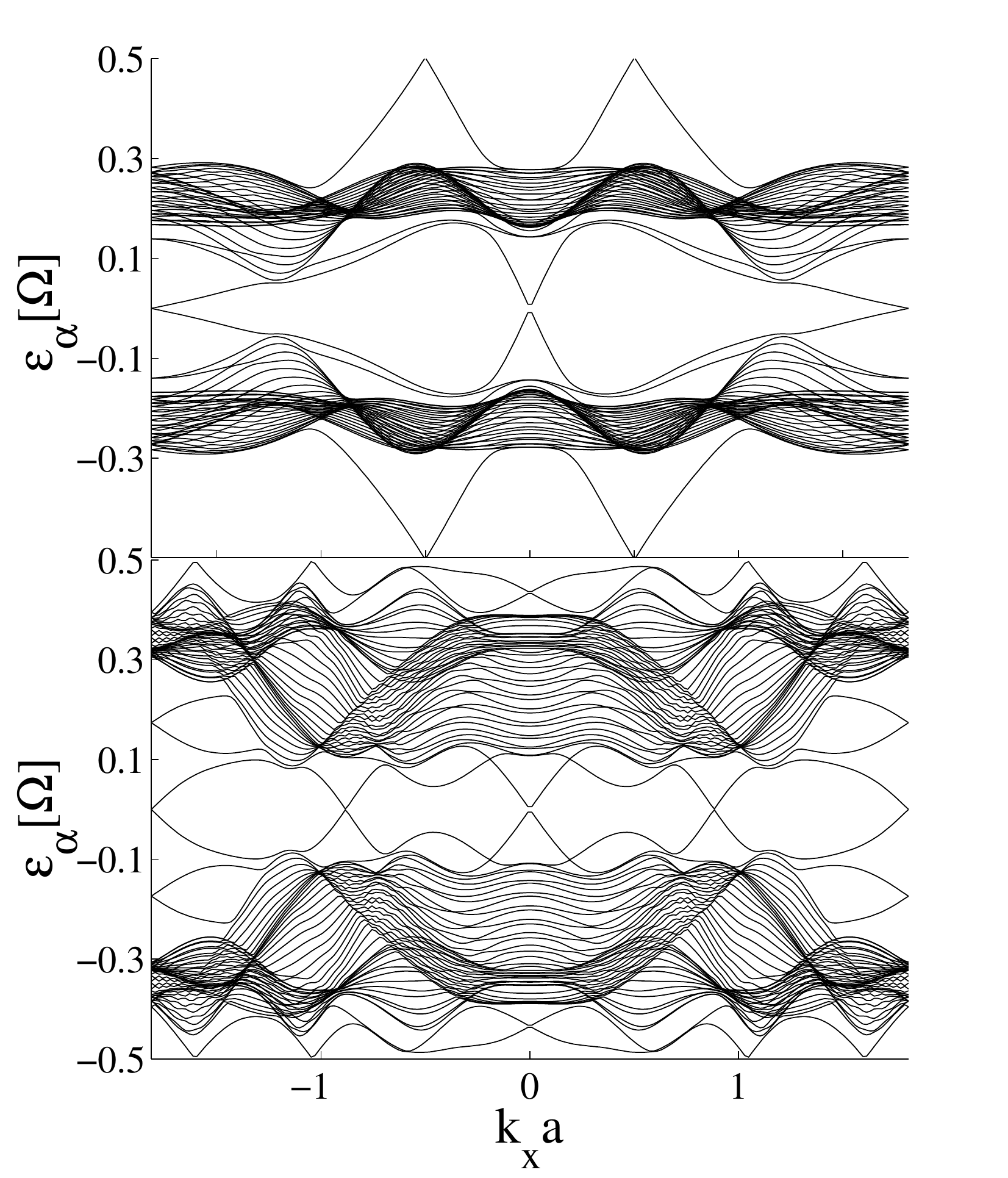}
\caption{Quasi-energies in a cylindrical geometry highlighting edge states. The laser frequency is
$\Omega=0.5t_h$, while the laser amplitude and Chern number are (upper panel) $A_0a=10, C=0$  and (lower panel) $A_0a=5,C=6.0$.
It is the top phase ($C=0$) whose Hall response is discussed in detail in the text.
}
\label{figedg2}
\end{figure}

\begin{figure}
\centering
\includegraphics[height=10cm,width=10cm,keepaspectratio]{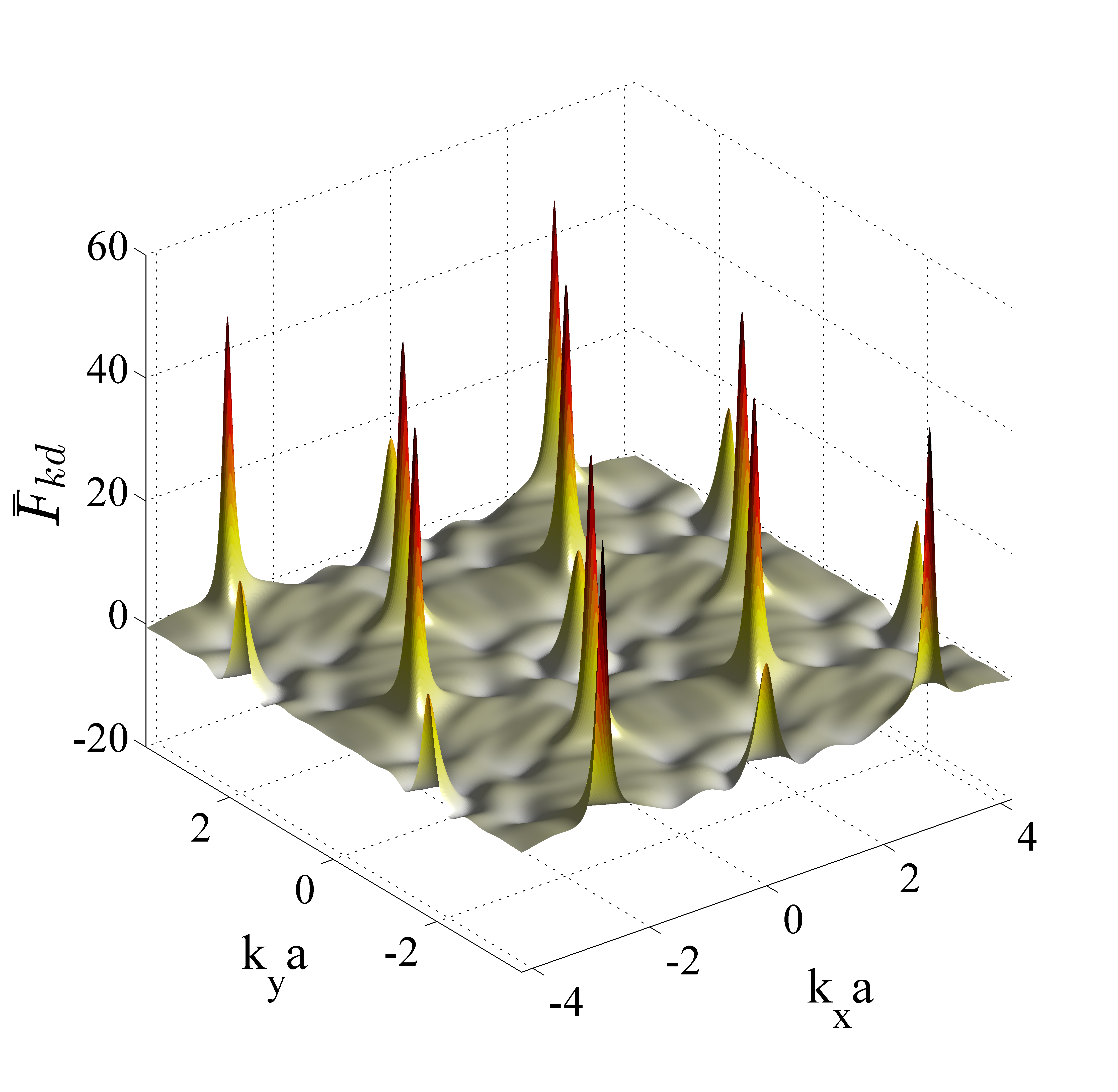}
\caption{(color online) Time-averaged Berry curvature over 4 BZs for $\Omega=0.5t_h,A_0a=10$ corresponding to
a Chern number of $C=0$.
}
\label{figFkA10}
\end{figure}

\begin{figure}
\centering
\includegraphics[height=8cm,width=9cm,keepaspectratio]{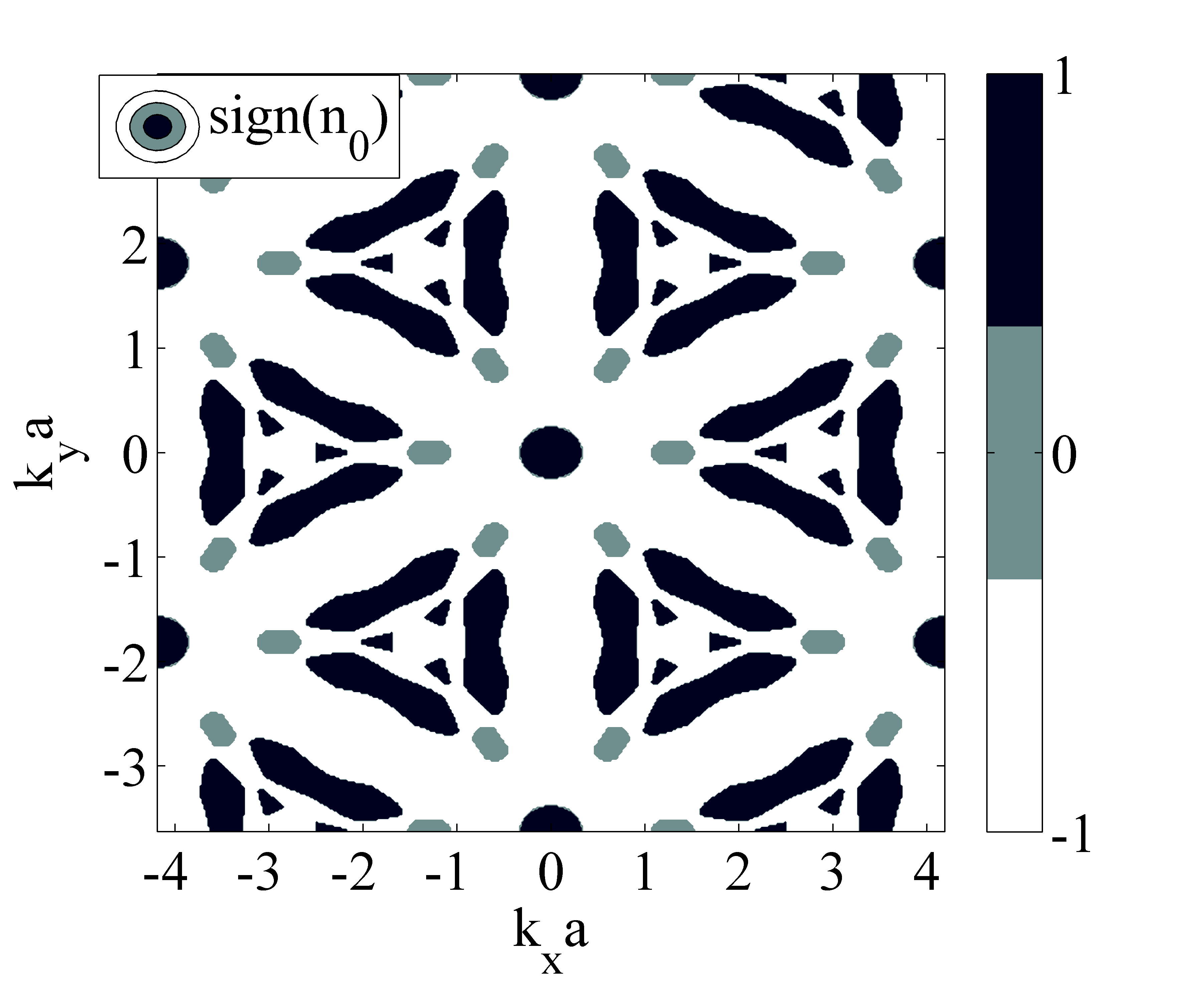}
\caption{(color online)
${\rm Sign}(n_0)$ in the dominant matrix element $M_k^{n_0}$ for $\Omega=0.5t_h,A_0a=10,C=0$. The sign of the population
imbalance ($\rho_{ku}-\rho_{kd}$) entering the Hall response follows this pattern.
}
\label{MknA10}
\end{figure}

One way to differentiate between this subtle matrix-element effect, and simply heating, where the latter will
also give a low $\sigma_{xy}$, is by studying the temperature dependence.
As the temperature is increased, one finds that the Hall conductivity actually increases (opposite to the heating case). This
happens because the quasi-energy level separation near the Dirac points
and around the circles is not the same, with the former being larger than the latter.
Raising the temperature excites electrons to the "down" level near the circles, while excitations to the
"up" level near the Dirac cones do not occur until higher temperatures. This results in a low temperature region where the Hall conductivity increases with temperature because in this regime as the temperature increases, the effective population imbalance 
between the two Floquet bands with opposite Berry curvature increases. This qualitative behavior is not special to these parameters, but holds for the entire topological phase where the only
difference is the size of the circle in Fig.~\ref{FkOm5}, and more generically is a signature of the fact that more than one scattering rate plays a dominant role in transport.

We now discuss this strong matrix element dependence in another phase of the FTI. We choose a much lower laser frequency, but larger
laser amplitude ($\Omega=0.5t_h, A_0a=10, C=0$) where one expects several band inversions as resonances involving
$n=1 \ldots 12$-photon processes are in principle allowed, although how visible these are, depends on the
laser amplitude. The corresponding quasi-energy spectra and Berry-curvature are shown in Fig.~\ref{figedg2} and~\ref{figFkA10} respectively.

For $A_0a=10$, $M_k^n$ has contributions from many $n$ where the dominant matrix element $M_k^{n_0}$ is such that
$-4 \lesssim n_0\lesssim 4$ in the BZ.
Fig.~\ref{MknA10} shows how the sign of $n_0$ varies through the BZ, where for $n_0=0$ or a positive
integer, the population is primarily in the "down" level at low temperatures, while $n_0$-negative gives a population primarily in the "up" level.
The sign of the
population difference follows the pattern in Fig.~\ref{MknA10}, leading to a dc Hall conductivity that is not only small, but also has a non-monotonic
temperature dependence. Changing the laser parameters in such a way as to stay within the same topological phase,
simply changes the size of the pockets of ${\rm max}(n_0)$
in the BZ, with no crossings or no new pockets appearing. Note that this particular phase
is not a usual Chern insulator as $C=0$, yet the system does support edge states. Moreover, while we are studying the system in the bulk,
and edges do not enter explicitly in our Kubo formula approach, yet we do pick up a non-zero Hall response coming from the
non-zero Berry-curvature of the bulk bands. 

\begin{figure}
\centering
\includegraphics[height=8cm,width=8cm,keepaspectratio]{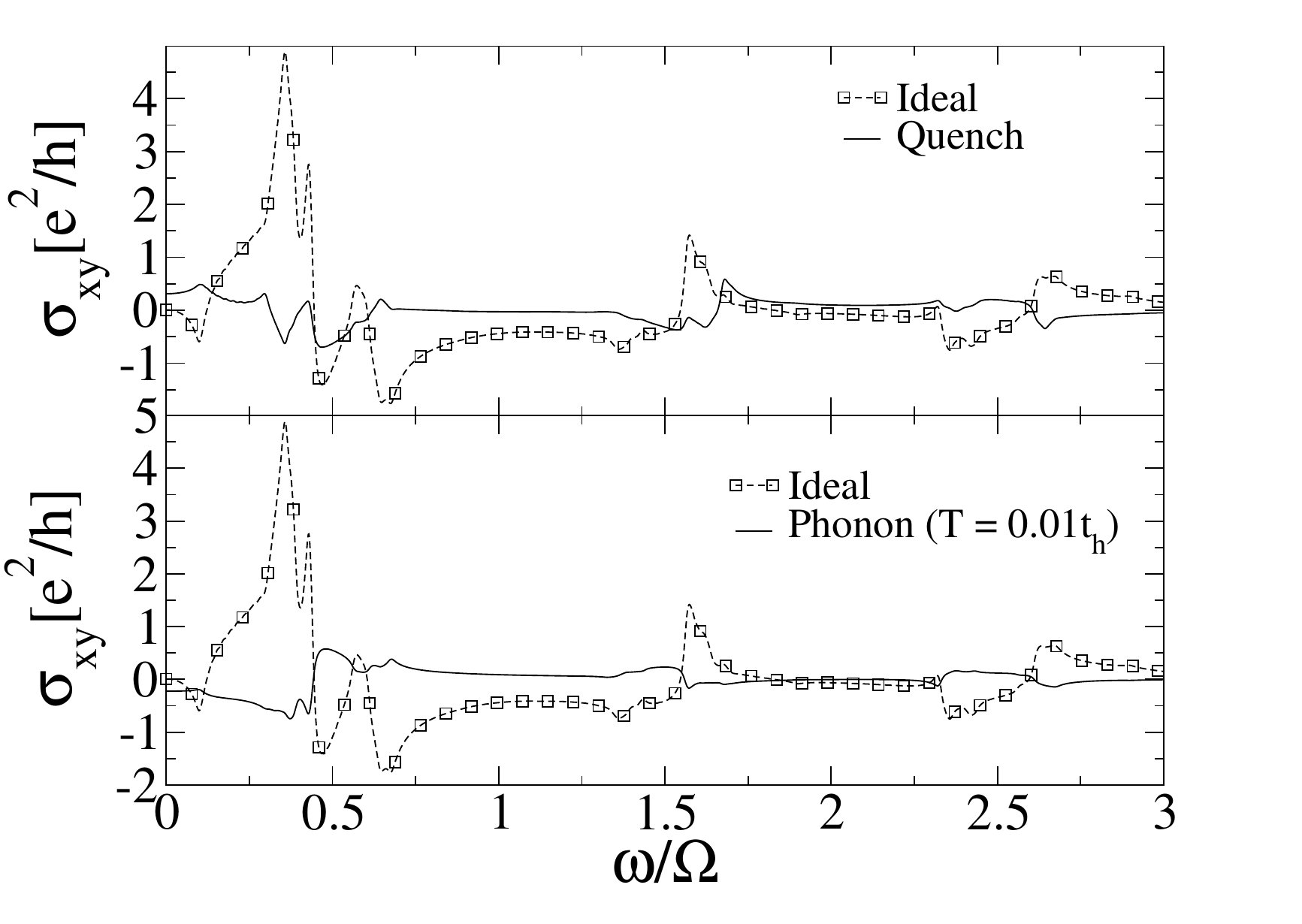}
\caption{$\sigma_{xy}(\omega)$ for $\Omega=0.5t_h,A_0a=10,C=0$. In comparison to the case of $\Omega=5.0t_h,A_0a=1.5,0.5$,
many side-bands are visible, and
the response is sensitive to the distribution function over a larger frequency range $\omega$.
}
\label{OptA10}
\end{figure}

\begin{figure}
\centering
\includegraphics[height=8cm,width=8cm,keepaspectratio]{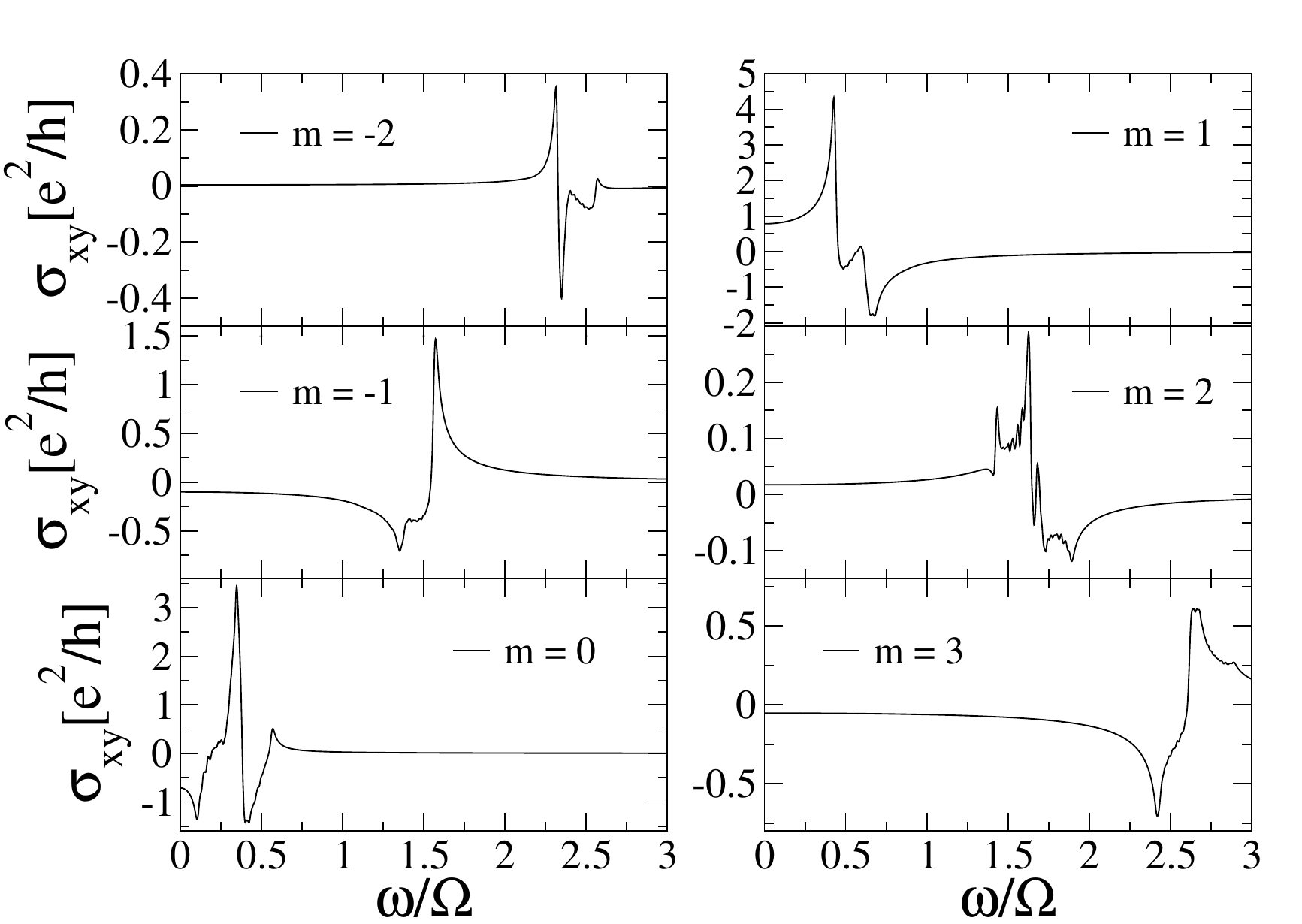}
\caption{$\sigma_{xy}^m$ for $\Omega=0.5t_h,A_0a=10,C=0$ and for an ideal distribution $\rho_{ku}-\rho_{kd}=1$. The peaks are
at $\omega \sim |\epsilon_{ku}-\epsilon_{kd}-m\Omega|$ where the location of $k$ is determined by the peaks in $F_k^m$.
}
\label{sigmamA10}
\end{figure}

While so far we have been mainly discussing the low frequency Hall response, we
now make some general observations about the high frequency response.
We find that for small amplitudes $A_0a\simeq {\cal O}(1)$, $\sigma_{xy}(\omega)$ is
saturated by $\omega\sim \Omega$ (Fig.~\ref{figOm5}). On the other hand, for larger amplitudes, like the case just discussed
($A_0a\simeq {\cal O}(10)$),
$\sigma_{xy}$ is non-zero over a larger range of frequencies (see Fig.~\ref{OptA10}), with the position and magnitude of the side-bands
depending on the structure of $F_k^m$. This is highlighted in Fig.~\ref{sigmamA10}  where the structure in the total optical Hall 
conductivity plotted in Fig.~\ref{OptA10} has been decomposed into those arising from separate Fourier components of $F_k$. 

To understand the location of the peaks in $\omega$, it is convenient to look again at the low amplitude case ($\Omega=5.0t_h,A_0a=1.5\&0.5$). 
Here only
$F_k^{m=0,1}$ are dominant, but while for one ($A_0a=1.5$), $F_{k}^{m=1}$ is peaked at $\epsilon_{ku}-\epsilon_{kd}\ll \Omega$, for
the other ($A_0a=0.5$), $F_k^{m=1}$ is peaked at $\epsilon_{ku}-\epsilon_{kd}\sim \Omega$. In fact $F_k^{m=0},F_k^{m=1}$ are 
respectively the Dirac cones and circles in Fig.~\ref{FkOm5}. Thus
a side-band appears at $\omega \sim |\epsilon_{ku}-\epsilon_{kd}-\Omega|\sim 0.75\Omega$ for the case of $A_0a=1.5,C=1$, while no such
side-band is visible for $A_0a=0.5,C=3$.
Similar arguments can be utilized to understand the peaks in the high frequency response for the large amplitude case shown in
Figs.~\ref{OptA10} and ~\ref{sigmamA10}. Thus while the optical Hall conductivity at non-zero frequency $\sigma_{xy}(\omega \neq 0)$
is not a topological quantity, yet it is sensitive to the structure of the Berry curvature both in momentum and time.

\section{Conclusions}\label{conclu}
In summary we have presented results for the optical Hall conductivity which can be measured in an all optical measurement such
as Faraday rotation, as well as using leads. The low frequency Hall conductivity, even in steady-state with a reservoir,
is remarkably sensitive to the topological phase, which is communicated through the structure of the system-reservoir matrix elements,
resulting
in steady-states with anomalous temperature dependence, and also opening up the possibility of manipulating the steady-state
by suitable reservoir engineering~\cite{Lindner15,Iadecola15}, for example by applying strain fields that modify the symmetries of the
electron-phonon coupling.
We also find that the high-frequency response acts as a spectroscopic tool for the Fourier components of the time-dependent
Berry curvature.

We presented detailed results for the optical Hall response in three different phases of the FTI. One is where the
Chern number coincided with the number of edge-states which were all 
in the center of the FBZ ($\Omega=5 t_h,A_0a=1.5,C=1$), the other where the Chern number does not equal the number
of edge-states but rather equals the difference in the number of chiral edge-states above and below the Floquet 
band ($\Omega=5 t_h,A_0a=0.5,C=3$), while the third was one where $C=0$ ($\Omega=0.5 t_h,A_0a=10,C=0$), 
so that the band was not a conventional Chern insulator, yet the
system supports edge states with the number of chiral edge modes above and below the band equal to each other (hence $C=0$).
An interesting future direction of research is to compare the results presented here which were 
based on a bulk Kubo formula treatment, and which
is sensitive to the bulk Berry-curvature, with a Landauer transport approach through a short sample where the leads determine the
occupation of the Floquet states.

{\sl Acknowledgments:}
The authors thank Andrea Cavalleri and Takashi Oka for helpful discussions.
This work was supported by the US Department of Energy,
Office of Science, Basic Energy Sciences, under Award No.~DE-SC0010821.


%

\begin{appendix}
\section{Derivation of Floquet-Master equation}\label{ap1}

We assume that initially the system is in the ground
state of graphene $|\Psi(t=0)\rangle=\prod_k|\psi_{\rm in,k}\rangle$, and we time-evolve
the system according to $H$. For the closed system ($H_c=0$), this time-evolution can
be studied exactly, while in the presence of phonons, we solve the
problem by assuming a weak electron-phonon coupling, and a very fast bath
so that a Floquet-Markov approximation can be made.
This approach has been discussed in detail elsewhere~\cite{Dehghani14,Dehghani15a}, however
for completeness we briefly highlight the main steps.
Let $W(t)$ be the full density matrix obeying
$\frac{dW(t)}{dt} = -i \left[H,W(t)\right]$.
In the interaction representation,
$W_{I}(t) = e^{i H_{\rm ph} t}U^{\dagger}_{\rm el}(t,0)W(t)U_{\rm el}(t,0)e^{- i H_{\rm ph} t}$,
where $U_{\rm el}(t,t')=\prod_kU_k(t,t')$ is the time-evolution operator
for the electrons under a periodic drive where
$U_k(t,t_0)=\sum_{\alpha=u,d}e^{-i\epsilon_{k\alpha}(t-t_0)}|\phi_{k\alpha}(t)\rangle\langle\phi_{k\alpha}(t_0)|$
with $|\phi_{k\alpha}(t)\rangle,\epsilon_{k\alpha}$ being the quasi-modes and quasi-energies that obey
$\biggl[H_{\rm el}-i\partial_t\biggr]|\phi_{k\alpha}\rangle = \epsilon_{k\alpha}|\phi_{k\alpha}\rangle$. We
will follow the convention of restricting the quasi-energies to a Floquet Brillouin zone (FBZ),
$-\Omega/2<\epsilon_{k\alpha}<\Omega/2$, and label the two quasi energy-levels such that $\epsilon_{ku}>\epsilon_{kd}$. Defining the electron reduced
density matrix as the one obtained from tracing over the phonons,
$W^{\rm el} = {\rm Tr}_{\rm ph}W$, at ${\cal O}(H_c^2)$ we need to solve,
\begin{eqnarray}
\frac{dW^{\rm el}_I}{dt}=-{\rm Tr}_{\rm ph}\int_{t_0}^tdt'\left[H_{c,I}(t),\left[H_{c,I}(t'),W_{I}(t')\right]\right]
\end{eqnarray}

We assume that at the initial time $t_0$, the electrons and phonons are uncoupled so that
$W(t_0) = W^{\rm el}_0(t_0)\otimes W^{\rm ph}(t_0)$, and that
initially the electrons are in the post-quench state $|\Psi(t)\rangle=\prod_kU_k(t,0)|\psi_{\rm in,k}\rangle$.
Thus,
$W^{\rm el}_0(t)= |\Psi(t)\rangle\langle \Psi(t)|=\prod_kW^{\rm el}_{k,0}$
where
\begin{eqnarray}
\!\!W^{\rm el}_{k,0}(t)=\!\!\! \sum_{\alpha,\beta=u,d}e^{-i(\epsilon_{k \alpha}-\epsilon_{k \beta})t}|\phi_{k\alpha}(t)\rangle
\langle\phi_{k\beta}(t)|
\rho_{k,\alpha\beta}^{\rm quench}
\end{eqnarray}
with
$\rho_{k,\alpha\beta}^{\rm quench}= \langle \phi_{k\alpha}(0)|\psi_{{\rm in},k} \rangle\langle\psi_{{\rm in},k}
| \phi_{k\beta}(0)\rangle$. Since the phonons are an ideal reservoir that stay in thermal equilibrium
at temperature $T$ at all times, we write $W_I(t) = W^{\rm el}_I(t)\otimes
e^{-H_{\rm ph}/T}/{\rm Tr}\left[ e^{-H_{\rm ph}/T}\right]$ (we set $k_B=1$).

The most general form of the reduced density matrix
for the electrons is
$W^{\rm el}_I(t)  =\prod_k \sum_{\alpha\beta}\rho_{k,\alpha \beta}(t) |\phi_{k\alpha}(t)\rangle\langle\phi_{k\beta}(t)|$
where in the absence of phonons, $\rho_{k,\alpha \beta}=\rho_{k,\alpha\beta}^{\rm quench }$ and are time-independent
in the interaction representation.
The last remaining step is to
make the following three approximations~\cite{Hanggi2005,Kohn01,Kohn09}: (a).
$\rho_{k,\alpha\beta}$
are slowly varying as compared to the characteristic time scales of the reservoir (Markov approximation), (b) they are
also slowly varying in comparison to the laser frequency
so that the scattering rates can be averaged over one cycle of the laser (modified rotating wave approximation)
and, (c) we are away from any topological transitions so that
the spacing between quasi-energy levels is
large as compared to the coupling to the bath ($|\epsilon_{ku}-\epsilon_{kd}-n\Omega|> D_{\rm ph}\lambda^2,|n\Omega|>D_{\rm ph}\lambda^2$),
under such conditions the
off-diagonal matrix elements ($\rho_{kud}$) are small and can be dropped.
These approximations lead to the rate equation,
$\dot{\rho}_{k,\alpha\alpha}(t)
=-\sum_{\beta=u,d}L^k_{\alpha\alpha;\beta\beta}
\rho_{k,\beta\beta}(t)\label{rate}$
where
$L^k_{\alpha\alpha,\beta\beta}$ are the scattering rates.

\end{appendix}

\end{document}